\documentclass{article}

\usepackage{arxiv}
\usepackage[utf8]{inputenc} 
\usepackage[T1]{fontenc} 
\usepackage{hyperref}       
\usepackage{url}            
\usepackage{graphicx}
\usepackage{graphics}
\usepackage{physics}
\newcommand{\beq}{\begin{equation}}
\newcommand{\eeq}{\end{equation}}
\usepackage{float}
\usepackage[mathscr]{eucal}
\usepackage{authblk}
\usepackage{titling}
\usepackage{amssymb} 
\newcommand{\subtitle}[1]{%
  \posttitle{%
    \par\end{center}
    \begin{center}\large#1\end{center}
    \vskip0.5em}%
}

\title{Lissajous coherent states via projection}

\author{
  \textbf{Errico J. Russo}$^1$, \textbf{James Schneeloch}$^2$, \textbf{Edwin E. Hach III}$^{1*}$, \\
  \textbf{Richard J. Birrittella}$^3$, \textbf{Wanda Vargas}$^4$, and \textbf{Christopher C. Gerry}$^4$ \\
  \vspace{2pt} \\ 
  $^1$School of Physics and Astronomy, Rochester Institute of Technology, \\
  Rochester, New York 14623, USA \\
  $^2$Air Force Research Laboratory, Information Directorate, \\
  Rome, New York 13441-4514, USA \\
  $^3$Booz Allen Hamilton, McLean, Virginia 22102, USA\\
  $^4$Department of Physics and Astronomy, Lehman College, \\
  The City University of New York, Bronx, New York 10468-1589, USA\\
  $^*$Author to whom any correspondence should be addressed.\\
  \texttt{eehsps@rit.edu}
}

\begin{document}
\maketitle

\begin{abstract}
	\noindent We construct stationary coherent states concentrated on Lissajous figures of the isotropic and anisotropic harmonic oscillators, the latter having coprime frequencies, by projecting products of ordinary coherent states (one coherent state for each degree of freedom) onto sets of degenerate states. By performing these projections, we are deriving our states from sets of coherent states that are known to follow the classical motion of a two-dimensional harmonic oscillator for arbitrary frequencies.  We clarify the nature of any singularities present in the phase of the wavefunction for each of the states we derive, and we establish a rigorous connection between the laminar flow of probability current and the emergence of quantum interference. Through this analysis, we are able to provide a clear and quantifiable definition for a vortex state of the two-dimensional harmonic oscillator (2DHO).  In an appendix, we show that our stationary states are true coherent states as they can be used to resolve the relevant identity operators (the above mentioned projection operators) on their respective degenerate subspaces. In the special case of the isotropic oscillator, the states obtained are the SU(2) coherent states, and we derive from our formalism the familiar resolution of unity for these states. 
\end{abstract}


\section{Introduction}
Over the years, several articles have appeared in the literature devoted to the appearance of Lissajous figures in the quantum mechanical treatment of the anisotropic two-dimensional harmonic oscillator (2DHO). See, for example, references \cite{chen_vortex_2003,makowski_comment_2005,gorska_correspondence_2006,kumar_commensurate_2008,doll_lissajous_2007}. Prior to that work, there appeared papers by De Bièvre \cite{bievre_oscillator_1992} and Pollet et al. \cite{pollett_elliptic_1995} on the connection between the wavefunctions and the classical periodic orbits for an isotropic 2DHO. The wave functions they found, which are constructed as superpositions of degenerate energy eigenstates, are concentrated on the classical elliptical orbits. It has long been known that the $n$-dimensional isotropic harmonic oscillator possesses as its degeneracy group (dynamical symmetry group) SU($n$) \cite{jauch_problem_1940,baker_degeneracy_1956,dulock_degeneracy_1965,loebl_symmetry_2014}, therefore it is perhaps not surprising that the wavefunctions of the isotropic 2DHO that are concentrated on the classical elliptical orbits are the SU(2) coherent states \cite{radcliffe_properties_1971,arecchi_atomic_1972} as defined over the Schwinger realization of the su(2) Lie algebra in terms of a pair of Bosonic annihilation and creation operators \cite{schwinger_angular_2015} (see below).  These SU(2) coherent states for isotropic 2DHO consist of a superposition of all the degenerate states of a given SU(2) multiplet. They may be obtained from the action the SU(2) displacement operator \cite{arecchi_atomic_1972} on an extremal state. We underscore that these SU(2) coherent states, because they are composed of degenerate energy eigenstates, do not evolve. They are stationary coherent states which are concentrated on classical elliptical orbits. SU(2) is the relevant group here and appears in this context as a dynamical symmetry as the Hamiltonian depends only on the SU(2) Casimir operator for this potential (see below). But there are potentially any number of states that can be formed as superpositions of the degenerate states of the isotropic 2DHO. 
What is so special about the SU(2) coherent states, if anything? We shall present an answer to this in Section 2.

Non-trivial Lissajous figures appear in the classical motion of an anisotropic 2DHO where the ratio of the two frequencies can be expressed as the ratio of coprime numbers \cite{pastana_using_2021,davis_classical_1986,symon_mechanics_1980}.  Several authors have discussed wave packets constructed from degenerate states associated with a given energy eigenvalue and have shown them to be concentrated on Lissajous-like figures of the corresponding classical orbits, but there does not seem to be a systematic method of specifying those states.  Chen and Huang \cite{chen_vortex_2003} have constructed such states by using the same coefficients as used for the isotropic case, i.e. they used the SU(2) coherent state coefficients. They even call their states “SU(2) coherent states.” This association was, in part, motivated by the fact that degeneracies in these anisotropic cases are also $N+1$-fold, though the dynamical group aspects of this are not too clear. (More on this below.) Otherwise, this choice of coefficients is not especially well-motivated. In fact, the construction of Chen and Huang, as pointed out by Kumar and Dutta-Roy \cite{kumar_commensurate_2008}, is really an ansatz. 
Kumar and Dutta-Roy \cite{kumar_commensurate_2008}, on the other hand, draw on the work of Louck et al. \cite{louck_canonical_1973} which involves a non-bijective canonical transformation within a degenerate eigenspace that maps a commensurate anisotropic oscillator onto an isotropic oscillator, which leads to a Schwinger realization of SU(2) involving canonically transformed annihilation and creation operators, which, in turn, leads to SU(2) coherent states formed from the degenerate energy eigenstates. However, these authors display no Lissajous figures.
As mentioned, the various approaches to states with wave functions concentrated on stationary classical-like orbits cited above have not been systematic. In this paper we present a systematic procedure for establishing such states and in the process construct new forms of coherent states. We start by noting obvious fact that the 2DHO, whether isotropic or anisotropic, is separable into two one-dimensional harmonic oscillators in both classical and quantum mechanics. In the latter case, taking the motions to be along the $x$ and $y$ directions, one can introduce a product of Glauber \cite{glauber_quantum_1963} coherent states $|\alpha\rangle_x|\beta\rangle_y$ where $\alpha$  and $\beta$  are arbitrary complex numbers. These states are known \cite{howard_coherent_1987} to follow the classical motions in time according to $|\alpha\rangle_x|\beta\rangle_y\rightarrow|\alpha e^{-i\omega_xt}\rangle_x|\beta e^{-i\omega_yt}\rangle_y$  such that the corresponding wave function is $\Psi_{\alpha,\beta}(x,y)=\langle x|\alpha e^{-i\omega_xt}\rangle\langle y|\beta e^{-i\omega_yt}\rangle$, the centroid of which follows the classical motion in accordance with Ehrenfest’s theorem, regardless of the values of the two angular frequencies. Of course, of special interest are the cases where the frequencies are relatively coprime, in which case there will be sets of degenerate states within the set of allowed energy eigenvalues for the 2DHO. Our approach is to construct the corresponding identity operators within each set of degenerate eigenstates, and then apply those operators as projection operators onto the product of coherent states $|\alpha\rangle_x|\beta\rangle_y$   The states that result will be stationary states whose wave functions will be concentrated on the corresponding classical orbits, namely, Lissajous figures. The central point here is that these wavefunctions are derived systematically from wave functions known to follow the classical motion. The states we obtain are not the same as those of Chen and Huang \cite{chen_vortex_2003}. We will refer to our states as Lissajous coherent states (LCS).

To set the stage for what follows, we briefly review the general case of a two-dimensional oscillator for a particle of mass $m$ with arbitrary angular frequencies   associated with  the (orthogonal) $x$ and $y$ directions, respectively. The Hamiltonian is 

\begin{equation}
    H=\frac{p_x^2+p_y^2}{2m}+\frac{m}{2}(\omega_x^2 x^2+\omega_y^2 y^2)
    \label{eq:arb2DHOH}
\end{equation}
which upon introducing the raising and lowering operators
\begin{equation}
    a_x=\left(\frac{m\omega_x}{2\hbar}\right)^{\frac{1}{2}}\left(x+\frac{ip_x}{m\omega_x}\right),  a_x^\dagger=\left(\frac{m\omega_x}{2\hbar}\right)^{\frac{1}{2}}\left(x-\frac{ip_x}{m\omega_x}\right)
    \label{eq:BosonopsX}
\end{equation}
and
\begin{equation}
    a_y=\left(\frac{m\omega_y}{2\hbar}\right)^{\frac{1}{2}}\left(y+\frac{ip_y}{m\omega_y}\right),  a_y^\dagger=\left(\frac{m\omega_y}{2\hbar}\right)^{\frac{1}{2}}\left(y-\frac{ip_y}{m\omega_y}\right)
    \label{eq:BosonopsY}
\end{equation}
satisfying the commutation relations $\left[a_i,a_j^\dagger\right]=\delta_{ij}$   with $\left[a_i,a_j\right]=0=\left[a_i^\dagger,a_j^\dagger\right]$  due to the fact that $\left[x,p_x\right]=i\hbar $ and $\left[y,p_y\right]=i\hbar$, takes the form
\begin{equation}
    H=\hbar\omega_x a_x^\dagger a +\hbar\omega_y a_y^\dagger a_y +\frac{\hbar}{2}\left(\omega_x+\omega_y\right).
    \label{eq:arb2DHOHBoson}
\end{equation}
The term $\hbar\left(\omega_x+\omega_y\right)/2$ is, of course, the combined zero-point energies.  Under unitary evolution driven by the Hamiltonian given by Eq. (\ref{eq:arb2DHOHBoson}), we have

\begin{equation}
    |\Psi_{\alpha,\beta}(t)\rangle=e^{-iHt/\hbar}|\alpha\rangle_x|\beta\rangle_y=e^{-i(\omega_x+\omega_y)t/2}|\alpha e^{-i\omega_xt}\rangle_x|\beta e^{-i\omega_yt}\rangle_y
    \label{eq:GlauberTevol}
\end{equation}
where the familiar Glauber coherent states [18] are given in their number bases in the next section. In what follows we ignore the irrelevant global phase factor $e^{-i(\omega_x+\omega_y)t/2}$.  The centroid of the wave packet

\begin{equation}
    \Psi_{\alpha,\beta}(x,y,t)=\langle xy|\Psi_{\alpha,\beta}(t)\rangle=\langle x|\alpha e^{-i\omega_xt}\rangle_x\langle y|\beta e^{-i\omega_yt}\rangle_y
    \label{eq:2DHOWFTevol}
\end{equation}

will follow the classical motion in the $x-y$ plane for any frequencies $\omega_x$  and   $\omega_y$.

As we shall discuss in detail later, the stationary states we derive come in two basic varieties, distinguished by whether the corresponding current density vanishes identically, resulting in a standing wave Lissajous coherent state, or whether the current density is non-zero but steady over some finite region, commonly referred to as a vortex state. Further, related to this characterization scheme for the stationary states we derive, we propose a reinterpretation of the important connection between probability current density and quantum interference. In fact, we find that the correct interpretation is essentially opposite to the one suggested in Ref. \cite{chen_vortex_2003}.

The balance of this paper is organized as follows: In Section 2 we consider the isotropic 2DHO from the above-mentioned point of view; Section 3 extends this analysis to the anisotropic 2DHO having frequencies whose ratios are irreducible rational numbers. In Section 4 we summarize our results. Appendix A provides the details of our demonstration that our LCS are overcomplete and resolve the relevant unit operators, i.e., the projection operators defined over the relevant degenerate states. Finally, in Appendix B we present an alternative, algebraic, method for obtaining our states.

\section{The isotropic oscillator}

The simplest case is the two-dimensional isotropic harmonic oscillator of mass $m$ for which $\omega_x = \omega_y =\omega$   and the Hamiltonian takes the form  

\begin{equation}
    H_{iso}=\hbar\omega\left(a_x^\dagger a_x +a_y^\dagger a_y+1\right) 
    \label{eq:Hiso}
\end{equation}
As mentioned above, this system has SU(2) as its dynamical symmetry group, also known as its degeneracy group. That this is so is fairly easy to show by the introduction of the Schwinger realization \cite{schwinger_angular_2015} of the corresponding su(2) Lie algebra given by 

\begin{equation}
    J_{1}=\frac{1}{2}\left(a_x^\dagger a_y +a_x a_y^\dagger\right ),J_{2}= \frac{1}{2i}\left(a_x^\dagger a_y -a_x a_y^\dagger\right ),J_{3}=\frac{1}{2}\left(a_x^\dagger a_x -a_y^{\dagger} a_y\right )
    \label{eq:Schwinger}
\end{equation}
satisfying the commutation relations $\left[J_{i},J_{j}\right]=i\epsilon_{ijk}J_{k}$. In addition we introduce the operator

\begin{equation}
    J_{0}=\left(a_x^\dagger a_x +a_y^{\dagger} a_y\right ),
    \label{eq:Jzero}
\end{equation}
which commutes with all the other elements of the algebra, i.e. $\left[J_{0},J_{i}\right]=0, i=1,2,3.$   Furthermore, the Casimir operator for this algebra is 

\begin{equation}
    J_{1}^2+J_{2}^2+J_{3}^2=J_{0}\left(J_{0}+1\right),
    \label{eq:Jsquared}
\end{equation}
which means that $J_{0}$ is essentially a Casimir operator. Then we note that our Hamiltonian can be written as 

\begin{equation}
    H_{iso}=\hbar\omega\left(2J_{0}+1\right),
    \label{eq:HisoJ0}
\end{equation}
where $2J_{0}=a_{x}^{\dagger}a_{x}+ a_{y}^{\dagger}a_{y}$   is the total number of quanta contained in the two independent degrees of freedom. Equation (\ref{eq:HisoJ0})   emphasizes the fact the Hamiltonian is a function of a Casimir operator, which is the characteristic feature of dynamical symmetry groups.  This means that 

\begin{equation}
    \left[H_{iso},f\left(J_{1},J_{2},J_{3}\right)\right]=0
    \label{eq:HisoCommutes}
\end{equation}
for any function $f\left(J_{1},J_{2},J_{3}\right)$  of the su(2) operators. The corresponding degenerate eigenstates for the given number of quanta $N$ are $|K\rangle_{x}| N-K\rangle_y$  for $K=0,1,...,N$  where $N=1,2,...,\infty$  and where    
\begin{equation}
    |K\rangle_x|N-K\rangle_{y}=\frac{\left(a_{x}^{\dagger}\right)^K\left(a_{y}^{\dagger}\right)^{N-K}}{\sqrt{K!\left(N-K\right)!}}|0\rangle_x|0\rangle_{y}.
    \label{eq:2DHOProjFock}
\end{equation}
Note that the degenerate states $|K\rangle_{x}|N-K\rangle_{y}$  satisfy the eigenvalue equation 
\begin{equation}
    \left(a_x^\dagger a_x +a_y^{\dagger} a_y\right )|K\rangle_x|N-K\rangle_{y}=N|K\rangle_x|N-K\rangle_{y},K=0,1,...N,
    \label{eq:isoeigen}
\end{equation}
where we note again that $a_{x}^{\dagger}a_{x}+ a_{y}^{\dagger}a_{y}=2J_{0}$.The $N+1$  -fold degeneracies of these states for a given value of $N=2J_{0}$  correspond precisely to the dimensions of the unitary irreducible representations of SU(2). The SU(2) “displacement” operator is given by $\text{exp}\left(zJ_{+}-z^{*}J_{-}\right)$   where $z$ is usually parameterized according to $z=e^{-i\phi}\theta/2$  where $\theta$ and $\phi$ are the usual angles on the Bloch sphere. The action of this “displacement” operator on the extremal state $|0\rangle_{x}|N\rangle_{y}$   results in  

\begin{equation}
    |\zeta,N\rangle=\text{exp}\left(zJ_{+}-z^{*}J_{-}\right)|0\rangle_{x}|N\rangle_{y}=\left(1+|\zeta|^2\right)^{-N/2}\sum_{K=0}^{N}\binom{N}{K}^{1/2}\zeta^K|K\rangle_{x}|N-K\rangle_{y},
    \label{eq:SU2CohState}
\end{equation}
where $\zeta=e^{-i\phi}\tan\left(\theta/2\right),$   which is exactly of the form of an SU(2) coherent state as would be expected for the su(2) Lie algebra expressed in terms of two sets of Boson operators \cite{buzek_generalized_1989}. 

In contrast to the above, we here start with a product of coherent states associated with the two directions $x$ and $y$ as given by their number state decompositions: 

\begin{equation}
    |\alpha\rangle_x|\beta\rangle_y=\left[e^{-|\alpha|^2/2}\sum_{m=0}^{\infty}\frac{\alpha^m}{\sqrt{m!}}|m\rangle_x\right]\left[e^{-|\beta|^2/2}\sum_{n=0}^{\infty}\frac{\beta^n}{\sqrt{n!}}|n\rangle_y\right]=e^{-|\alpha|^2/2}e^{-|\beta|^2/2}\sum_{m=0}^{\infty}\sum_{n=0}^{\infty}\frac{\alpha^m\beta^n}{\sqrt{m!n!}}|m\rangle_x|n\rangle_y,
    \label{eq:Glauber}
\end{equation}
and then perform projections onto specific degenerate subspaces. For the case of the isotropic oscillator, the relevant projection operator onto the subspace of $N$ total quanta is 

\begin{equation}
    \Pi_N=\sum_{K=0}^N|K\rangle_x|N-K\rangle_{yx}\langle K|_y\langle N-K|.
    \label{eq:isoProj}
\end{equation}
Applying this operator to the product state $|\alpha\rangle_x|\beta\rangle_y$  we project out  

\begin{equation}
    \Pi_N|\alpha\rangle_x|\beta\rangle_y\propto\sum_{K=0}^N\frac{\alpha^K\beta^{N-K}}{\sqrt{K!\left(N-K\right)!}}|K\rangle_x|N-K\rangle_y,
    \label{eq:isoProjectionprop}
\end{equation}
which when normalized becomes

\begin{equation}
    |\zeta,N\rangle=\left(1+|\zeta|^2\right)^{-N/2}\sum_{K=0}^N\binom{N}{K}^{1/2}\zeta^K|K\rangle_x|N-K\rangle_y,
    \label{eq:NormSU2CS}
\end{equation}
where $\zeta=\alpha/\beta$.   This is exactly the same state as in Eq. (\ref{eq:SU2CohState}), but this time obtained without using group-theoretical methods. (In Appendix B we present yet another algebraic method for obtaining these states.) Furthermore, we can now answer the question raised in Section I about why the SU(2) coherent state is so special in this context. The answer is that it can be understood as being obtained from a product of classical-like states (the Glauber states) by a simple projection onto the degenerate subspaces.  

The configuration space wave functions associated with our states are given by 

\begin{equation}
    \Psi_N\left(x,y,\zeta\right)=\left(1+|\zeta|^2\right)^{-N/2}\sum_{K=0}^{N} \binom{N}{K}^{1/2}
    \zeta^K\psi_K\left(x\right)\psi_{N-K}\left(y\right),
    \label{eq:isoLJwf}
\end{equation}
where

\begin{equation}
    \psi_K\left(x\right)=\frac{1}{\sqrt{2^KK!}}\left(\frac{m\omega}{\pi\hbar}\right)^{1/4}\exp{\left[-\frac{m\omega}{2\hbar}x^2\right]}H_K\left(x\sqrt{m\omega/\hbar}\right)
        \label{eq:ODHOwfsK}
\end{equation}
and
\begin{equation}
    \psi_{N-K}\left(y\right)=\frac{1}{\sqrt{2^{N-K}\left(N-K\right)!}}\left(\frac{m\omega}{\pi\hbar}\right)^{1/4}\exp{\left[-\frac{m\omega}{2\hbar}y^2\right]}H_{N-K}\left(y\sqrt{m\omega/\hbar}\right)
    \label{eq:ODHOwfsN-K}
\end{equation}
and where the $H_K\left(x\sqrt{m\omega/\hbar}\right)$  etc. are, of course, the Hermite polynomials. 

To emphasize the strong connection between our states and the corresponding classical Lissajous figures, we consider the configuration space probability densities

\begin{equation}
    \rho_N\left(x,y,\zeta\right)=|\Psi_N\left(x,y,\zeta\right)|^2
    \label{eq:isoprobdens}
\end{equation}
along with the associated probability current densities 

\begin{equation}
    \vec{J}_N\left(x,y,\zeta\right)=\frac{\hbar}{m}\text{Im}{\left[\Psi_N^*\left(x,y,\zeta\right)\vec{\nabla}\Psi_N\left(x,y,\zeta\right)\right]}
    \label{eq:isocurrdens}
\end{equation}
for each case of interest.  

Local conservation of probability assumes the form of a continuity equation, 

\begin{equation}
    \vec{\nabla}\cdot\vec{J}_N\left(x,y,\zeta\right)+\frac{\partial\rho_N\left(x,y,\zeta\right)}{\partial t}=0
    \label{eq:continuity}
\end{equation}
which, in light of their stationary nature, constrains the states we consider to be ones having steady probability current densities, 

\begin{equation}
    \vec{\nabla}\cdot\vec{J}_N\left(x,y,\zeta\right)=0.
    \label{eq:steadycurr}
\end{equation}
Equation (\ref{eq:steadycurr}) allows us to define two distinct behaviors for stationary Lissajous coherent states via projection; the current density can vanish identically resulting in a standing wave state that manifests maximal quantum interference in the probability density, or the current density can exhibit non-zero, laminar flow of probability resulting in what is commonly and generically referred to as a vortex state exhibiting less quantum interference and in some cases none at all \cite{chen_vortex_2003}. 

\begin{figure}
\centering
       \includegraphics[width=1.0\textwidth]{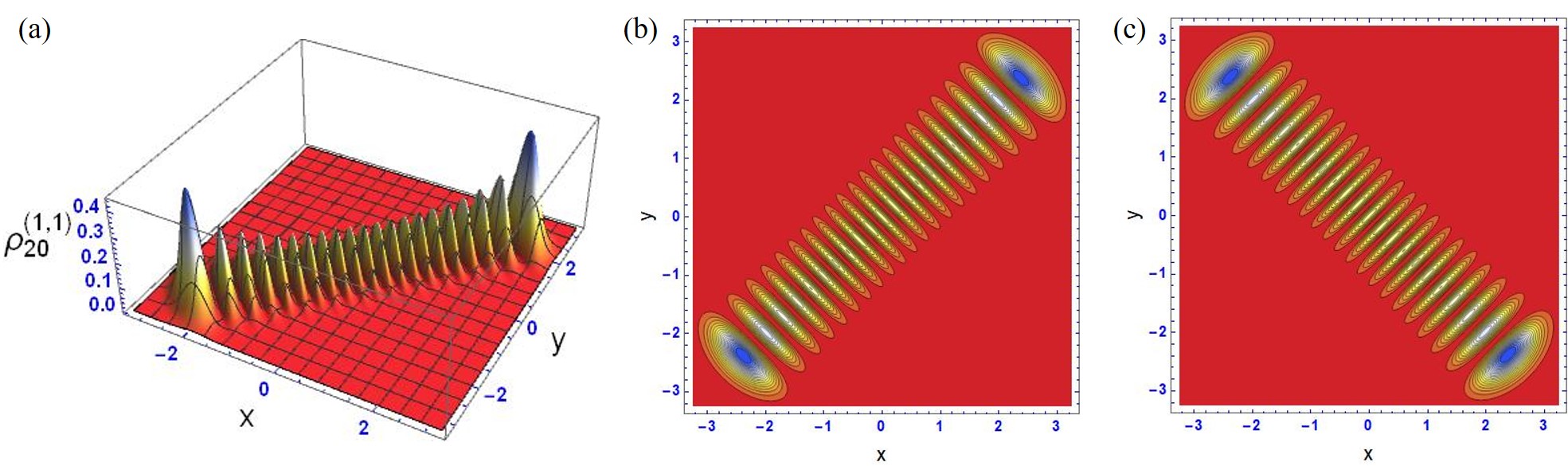}
 \caption{Probability density plots for the two independent standing wave Lissajous coherent states for the isotropic two-dimensional harmonic oscillator. Parts (a) and (b) show the surface and contour plots, respectively, for states satisfying the condition stated in Eq. (\ref{realwfphasecond}) with $n=0,\pm 2,\pm 4,...$, and Part (c) shows the contour plot for states having $n=\pm 1,\pm 3,\pm 5, ...$. For each of these cases the probability current density vanishes identically and, as discussed in the text, the interference fringes are maximally visible.}
\label{fig1:IsoStandingWave}
\end{figure}

To clarify the relationship between the phase of the complex amplitude and the probability current density for the state we follow the approach of Chen and Huang by expressing the configuration space wave function given in Eq. (\ref{eq:isoLJwf}) in polar form \cite{chen_vortex_2003}

\begin{equation}
    \Psi_N\left(x,y,\zeta\right)=\sqrt{\rho_N\left(x,y,\zeta\right)}\exp{\left[i\chi_N\left(x,y,\zeta\right)\right]}
    \label{isopolarwf}
\end{equation}
where

\begin{equation}
    \chi_N\left(x,y,\zeta\right)=\arctan\left[\frac{\text{Im}\left(\Psi_N\left(x,y,\zeta\right)\right)}{\text{Re}\left(\Psi_N\left(x,y,\zeta\right)\right)}\right]
    \label{isophase}
\end{equation}
Recalling the polar form for $\zeta$, it follows from Eq. (\ref{isophase}) that the wavefunction inherits its phase directly from the phase of the complex amplitude, $\zeta$, for the state vector. Notably, the wavefunction is real whenever $\zeta$  is real, which happens in the states given by Eq. (\ref{eq:NormSU2CS}) whenever 

\begin{equation}
    e^{i\phi}=\pm 1\implies\phi=n\pi, n=0,\pm1,\pm2,...
    \label{realwfphasecond}
\end{equation}
Enforcing Eq. (\ref{realwfphasecond}) results in the condition  

\begin{equation}
    \text{Im}\left(\Psi_N\left(x,y,\zeta\right)\right)=0
    \label{realwf}
\end{equation}
which, in light of Eq. (\ref{isophase}) ensures that the probability current density for any such state vanishes identically, as can be seen from Eq. (\ref{eq:isocurrdens}), resulting in a \textit{standing wave} Lissajous coherent state. In Fig. (\ref{fig1:IsoStandingWave}) we plot the probability density functions the two independent forms of the standing wave Lissajous coherent state for the two-dimensional isotropic oscillator. Aside from the obvious localization of the state along the corresponding classical Lissajous figure, the most important features evident in this figure are the sharply visible interference fringes resulting from the highly oscillatory nature of the standing wave probability density. In concert with this, an important feature not shown in Fig. (\ref{fig1:IsoStandingWave}) is the total cancellation of the probability current density; the standing wave states maximize quantum interference.

\begin{figure}
\centering
       \includegraphics[width=1.0\textwidth]{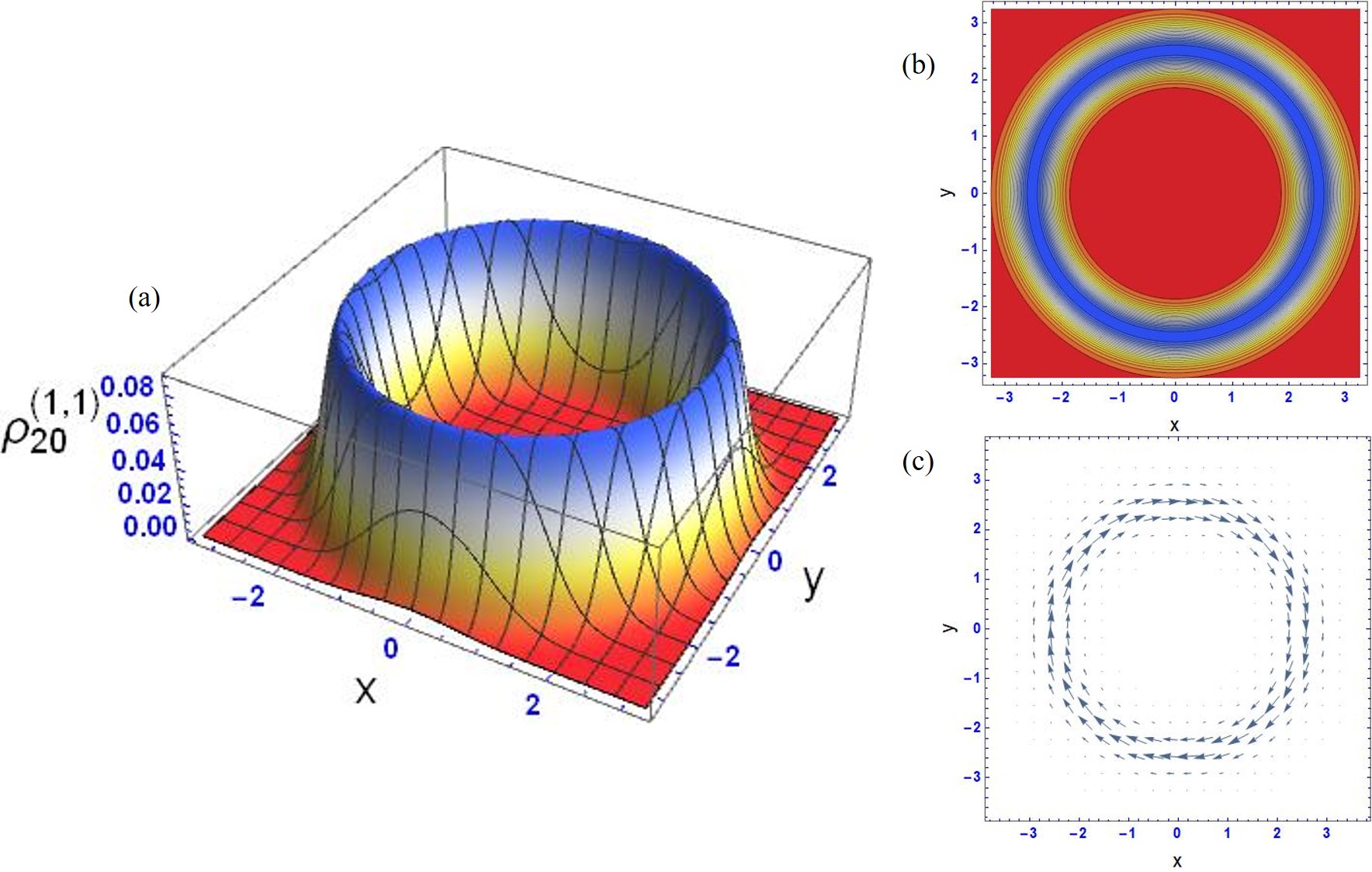}
 \caption{Probability, (a) and (b), and probability current, (c), density plots for a Lissajous Coherent State at the so-called \textit{vortex limit} for the isotropic two-dimensional harmonic oscillator.}
\label{fig1:IsoVortexLim}
\end{figure}

The opposite extreme case of a stationary LCS occurs for the isotropic oscillator whenever

\begin{equation}
    e^{i\phi}=\pm i\implies\phi=\left(n+\frac{1}{2}\right)\pi, n=0,\pm1,\pm2,...,
    \label{eq:vortexphase}
\end{equation}
resulting in a state for which $\text{Im}\Psi_N\left(x,y,\zeta\right)\neq0$ such that 

\begin{equation}
    \vec{\nabla}\cdot\vec{J}=0 \text{ with }\vec{J}\neq 0.
    \label{eq:vortcurr}
\end{equation}
Equation (\ref{eq:vortcurr}) expresses the general condition satisfied by a so-called \textit{vortex} state of the system \cite{chen_vortex_2003}. A vortex state in this context is one that featured steady, laminar flow of the probability current density generally localized along the corresponding classical Lissajous figure. We refer to any Lissajous coherent state for which the condition given in Eq. (\ref{eq:vortexphase}) holds as a state the vortex limit for the system; this is a state for which the circulation of probability current density is strongest. In Fig.(\ref{fig1:IsoVortexLim}) we plot probability densities and probability current densities for the two independent forms of Lissajous coherent states at the vortex limit for the isotropic two-dimensional oscillator. Whereas parts (a) and (b) of Fig. (\ref{fig1:IsoVortexLim}) are clearly identical for all of the states that satisfy Eq. (\ref{eq:vortexphase}), the probability current density flows clockwise, as shown in Fig. (\ref{fig1:IsoVortexLim}c), for $n$ even in Eq. (\ref{eq:vortexphase}) and counterclockwise for $n$ odd, the latter case not shown in Fig. (\ref{fig1:IsoVortexLim}).    

From Eq. (\ref{isopolarwf}), it is well known that one can compute the probability current density using the alternate form \cite{sakurai_modern_2017} 

\begin{equation}
\vec{J}\left(x,y,\zeta\right)=\frac{\hbar}{m}\rho_{N}\left(x,y,\zeta\right)\vec{\nabla}\chi_{N}\left(x,y,\zeta\right)
\label{eq:currentfromphase}
\end{equation}
from which it is apparent that standing wave states are represented by wavefunctions having phases that are independent of position. The situation is more interesting in regards to vortex states which require the steady circulation of a non-zero probability current density. For the moment, we consider the vortex limit state for the isotropic oscillator as depicted in Fig. (\ref{fig1:IsoVortexLim}).

\begin{figure}
\centering
       \includegraphics[width=1.0\textwidth]{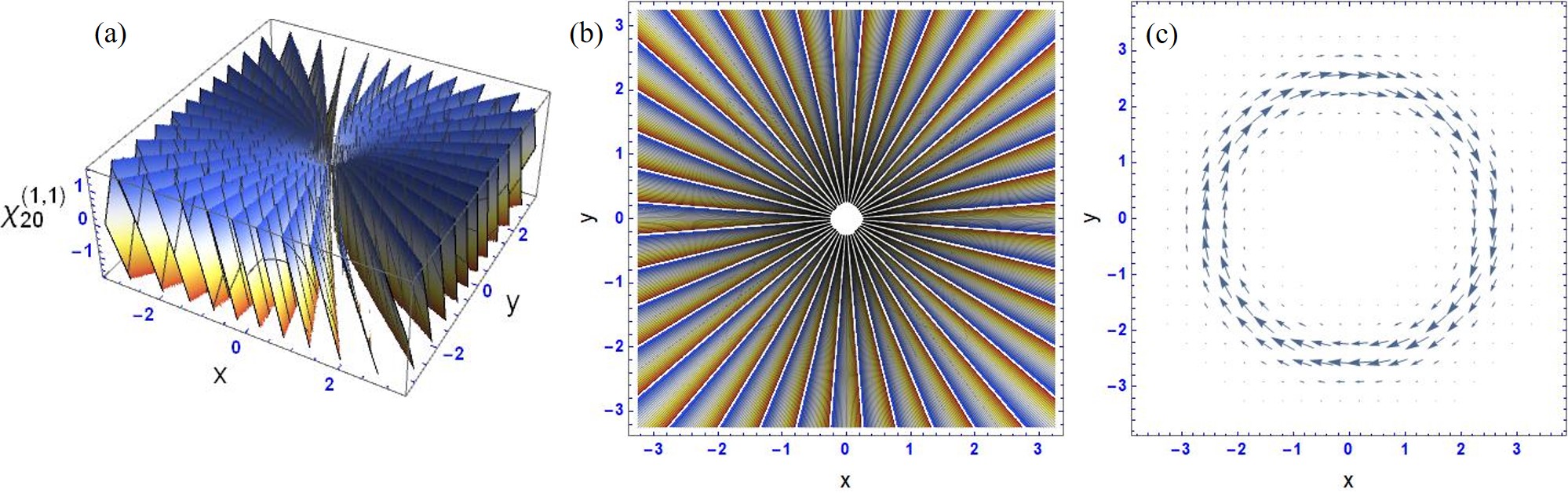}
 \caption{Surface (a) and contour (b) plots for the phase of the wavefunction $\chi\left(x,y\right)$ for $N=20$ of the vortex limit for Lissajous coherent states of the 2DHO along with the corresponding steady probability current density (c). The white 'dot' at the origin of (b) represents the non-physical, essential singularity in the phase at this point. The white rays extending radially from the essential singularity are jump discontinuities in the phase that are required to ensure the steady nature of the circulating probability current. The vertical axis of (a) gives the phase of the wavefunction $\mod2\pi$ between $-\pi$ and $\pi$.  }
\label{fig3:IsoVortexLWFphase}
\end{figure}

In Fig. (\ref{fig3:IsoVortexLWFphase}) we display the (a) surface and (b) contour plot for the phase of the wavefunction along with (c) the probability current density for the state at the vortex limit of the isotropic oscillator. 
The results shown in Fig. (\ref{fig3:IsoVortexLWFphase}) are in agreement with similar results found through other means by the authors of Ref. \cite{chen_vortex_2003}. Those authors refer correctly, but in a rather unspecific way to the \textit{singularities} in the phase of the wavefunction for this type of state. With our result in view, we now clarify the point. Referring to Fig. (\ref{fig3:IsoVortexLWFphase}), we see that there are two types of \textit{apparent} singularities present in the phase of the wavefunction. One of these is the trivial singularity at the center of circulation of the current density, located in this case in the center of the white dot in Fig. (\ref{fig3:IsoVortexLWFphase}a). The singularity at this point is of no physical consequence, and it is directly analogous to the ambiguity encountered in attempting to assign a longitude coordinate to a point at the North Pole of Earth. The other apparent singularity actually appears to be a jump discontinuity in the phase of the wavefunction occurring along the white rays extending radially outward from the trivial singularity in Fig. (\ref{fig3:IsoVortexLWFphase}b). These turn out not to be singularities of any kind, but, rather, mathematical artifacts of our choice to plot the phase$\text{ mod } 2\pi$. This is especially easy to see in the relatively simple geometry of the probability current density and phase of the wavefunction at the vortex limit for the isotropic 2DHO. Let $\hat{\varphi}$ be a clockwise unit vector along a streamline at radius $r$ in the probability current density shown in Fig. (\ref{fig3:IsoVortexLWFphase}c). The current density in this case is given by
\begin{equation}
\vec{J}\left(r,\zeta\right)=\frac{\hbar}{m}\rho_{N}\left(r,\zeta\right)\vec{\nabla}\chi_{N}\left(r,\varphi,\zeta\right),
    \label{eq:vortexlimitJ}
\end{equation}
which, in this case, is independent of $\varphi$, so that
\begin{equation}
\vec{J}\left(r,\zeta\right)\cdot\hat\varphi=\frac{\hbar}{m}\frac{\rho_{N}\left(r,\zeta\right)}{r}\frac{\partial}{\partial\varphi}\chi_{N}\left(r,\varphi,\zeta\right)=\left|\vec{J}\left(r,\zeta\right)\right|,
    \label{eq:vortexlimitJphi}
\end{equation}
implying that $\cfrac{\partial\chi_{N}}{\partial\varphi}$ is independent of $\varphi$ so that the phase of the wavefunction has the form
\begin{equation}
\chi_{N}\left(r,\varphi,\zeta\right)=F\left(r,\zeta\right)\varphi.
    \label{eq:phaselinearform}
\end{equation}
Equation (\ref{eq:phaselinearform}) is a constraint on the phase of the wavefunction induced ultimately by the requirement that the probability current density be steady for the stationary LCS, and it is to this constraint that we can trace the origin of the apparent jump discontinuities in Fig. (\ref{fig3:IsoVortexLWFphase}). The phase of the wavefunction is a linear function of $\varphi$ at all points on a circle of a given radius, but we are expressing that phase in the interval $\left[-1,1\right]\text{ mod } 2\pi$; the jumps merely accommodate the necessary resets at the endpoints of this interval. For an isotropic LCS defined on an $N+1$-fold degenerate subspace, the phase of the wavefunction mod $2\pi$ will cycle through this interval $2N$ times along a circle of any fixed radius. In fact, if we were to plot the phase shown in Fig. (\ref{fig3:IsoVortexLWFphase}a) in the range $[-20,20]$ the result would be a continuous helical surface; no jumps required. Doing so in Fig. (\ref{fig3:IsoVortexLWFphase}b) would result in the disappearance of the white radial rays. Illusionary jump discontinuities notwithstanding, the mod $2\pi$ plots are far more useful for understanding properties of the vortex states than are plots explicitly preserving the underlying continuity of the phase.  

\begin{figure}
\centering
       \includegraphics[width=1.0\textwidth]{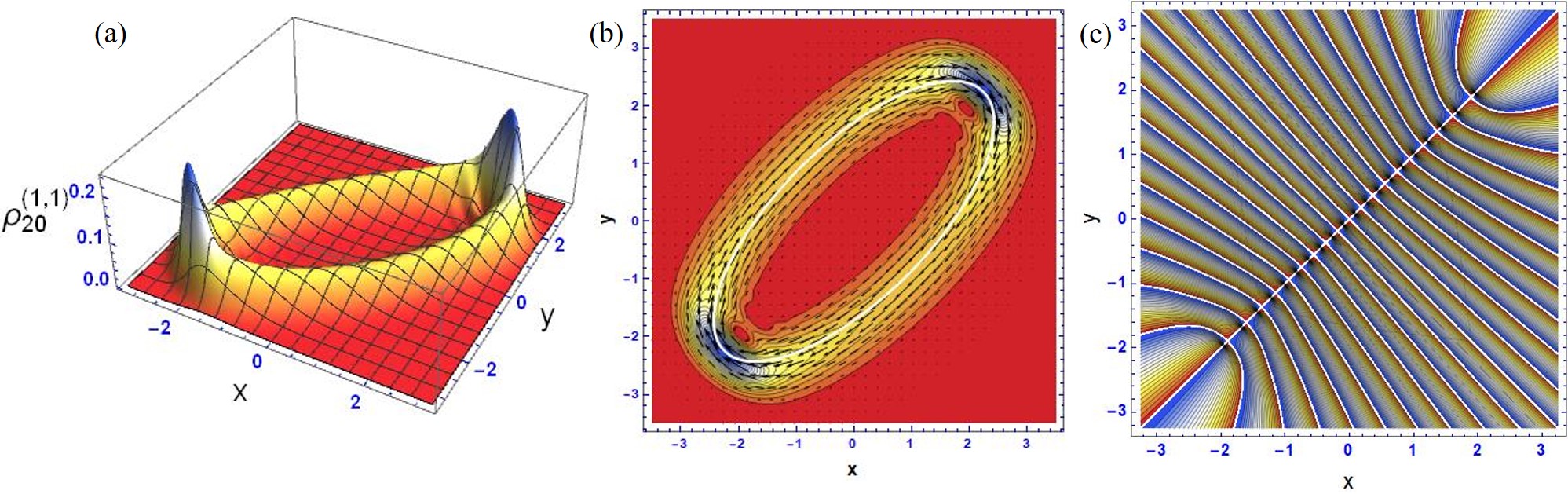}
 \caption{An intermediate vortex state for the isotropic 2DHO. Here the phase of the complex amplitude $\zeta$ is chosen to be midway between the condition for a standing wave state and that for the vortex limit as given by Eqs. (\ref{realwfphasecond}) and (\ref{eq:vortexphase}) (a) Surface plot of the probability density. (b) Contour plot of the probability density with the probability current density inlaid to emphasize the localization of each around the corresponding classical Lissajous figure (white ellipse). (c) Contour plot of the phase of the wavefunction $\chi\left(x,y\right)$ for $N=20$. The trivial essential singularity has stretched into a line segment along the major axis of the elliptical support for the probability and probability current densities. As in the case of the vortex limit, the rays of apparent jump discontinuities are simply artifacts of our restriction to the interval $[-\pi,\pi]\mod 2\pi$.  }
\label{fig4:IsoVortexIntermediate}
\end{figure}

So far, we have examined the phase of the wavefunction and its relationship with the probability current density only for the state at the vortex limit for the isotropic 2DHO. Although the geometry is a little more complex for other vortex states, our observations carry over directly. To illustrate this idea we show in Fig. (\ref{fig4:IsoVortexIntermediate}) the probability density, probability current density, and phase of the wavefunction for an \textit{intermediate} vortex state of the isotropic 2DHO. In particular, we have chosen the state for which the phase of the complex amplitude $\zeta$ is $\pi/4$, midway between the conditions for the vortex and standing wave limits. It is immediately apparent that the state is localized along the elliptical classical Lissajous trajectory for equal amplitude oscillators having the same relative phase shift. Furthermore, owing to the elliptical geometry laminar flow of the probability current, the trivial, essential singularity arising from the ambiguous nature of the flow along the major axis has morphed into a line segment in contrast with the singular point in the previous example. Once again, in Fig. (\ref{fig4:IsoVortexIntermediate}c) we see the artificial jump discontinuities that arise purely from our restriction of the range over which we plot the phase. The physical reality is still governed by Eq. (\ref{eq:currentfromphase}).  

\begin{figure}
\centering
       \includegraphics[width=1.0\textwidth]{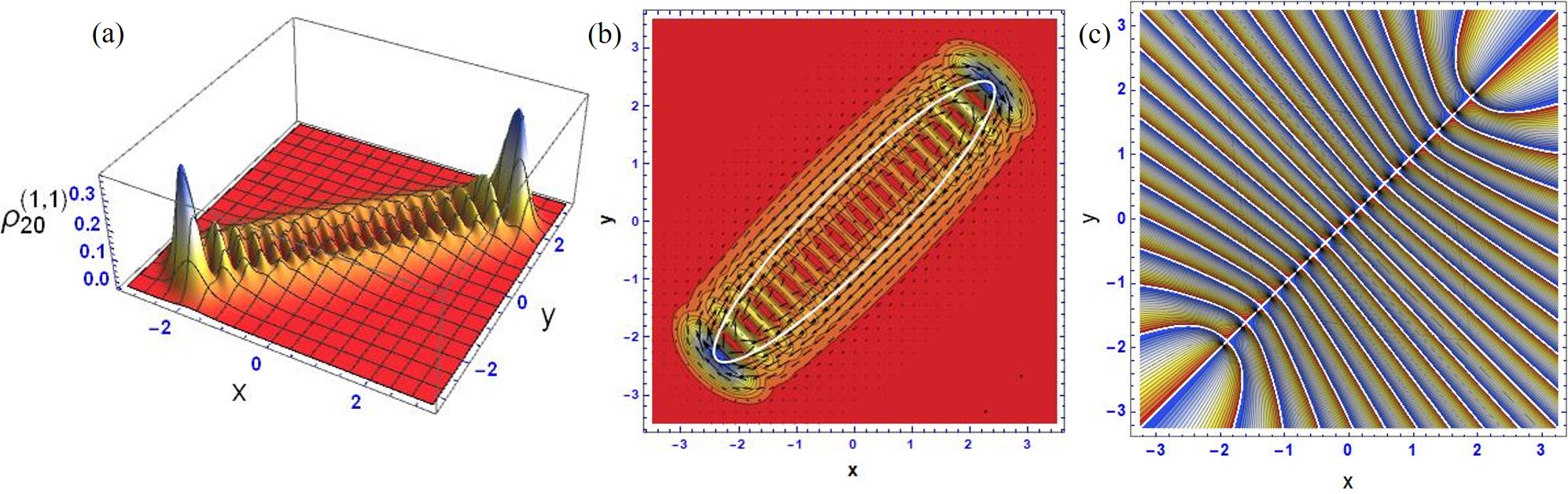}
 \caption{Another intermediate vortex state for the isotropic 2DHO, this one close to the standing wave limit than the one in Fig.(\ref{fig4:IsoVortexIntermediate}). Specifically, here the phase of the complex amplitude $\zeta$ is chosen to be $\pi/8$. Parts (a), (b), and (c) are directly analogous to the same parts of Fig.(\ref{fig4:IsoVortexIntermediate}) for the new choice of complex amplitude.  }
\label{fig5:IsoVortex_Pi_over_8}
\end{figure}

To complete our analysis of the LCS obtained via projection for the isotropic 2DHO, we plot in Fig. (\ref{fig5:IsoVortex_Pi_over_8}) the probability and probability current densities along with the phase of the wavefunction for the case in which the phase of the complex amplitude is $\pi/8$. Once again, the localization of the state along the corresponding classical Lissajous figure is evident. Furthermore the relationship between the laminar flow of the probability current density and the structure of the phase of the wavefunction progresses predictably into a more highly eccentric elliptical pattern, exactly as we expect from the previous discussion. Here, however, we see a new feature in the physics of the LCS. Referring to Fig.(\ref{fig5:IsoVortex_Pi_over_8}) we see the emergence of interference fringes in the region near and along the major axis of the Lissajous ellipse. 

In order to understand the basic mechanism underlying the onset of quantum interference in the LCS, it is helpful to consider the states represented in Figs. (\ref{fig1:IsoVortexLim}), (\ref{fig4:IsoVortexIntermediate}), (\ref{fig5:IsoVortex_Pi_over_8}), and (\ref{fig1:IsoStandingWave}), in that order, which is a progression of states starting at the vortex limit, where there are no interference fringes in the isotropic case, and ending at the standing wave limit, where quantum interference reaches its maximal sharpness and there is no probability current. Recalling that it is a defining property of the LCS that the support of the probability current density for such a state is non-negligible over a finite region centered upon the corresponding classical Lissajous trajectory, it is clear that counter-flowing portions of the current density are farther separated in lower eccentricity states closer to the vortex limit, see Fig. (\ref{fig1:IsoVortexLim}c), than in higher eccentricity ones closer to the standing wave limit, see Fig. (\ref{fig5:IsoVortex_Pi_over_8}b), resulting, in fact, in the complete cancellation of the current density at the standing wave limit. As a direct result of our choice of $\left|\zeta\right|=1$ for the complex amplitude of the state vector, the major axis of each LCS ellipse we have considered for the isotropic 2DHO falls along the line $y=x$. To clarify the argument, we can express the state vector for a LCS in the un-normalized form $|\zeta,N\rangle\sim|\zeta,N;+\rangle+|\zeta,N;-\rangle$, where the additional label $\pm$ indicates the branch of the state vector for which $\vec{J}\left(x,y\right)\cdot\left(\hat{i}+\hat{j}\right)/\sqrt{2}\gtrless 0$. In states near the vortex limit, where appreciable current flowing in opposite directions is well separated in space, these two branches of the state vector are approximately orthogonal, $\langle\zeta,N;+|\zeta,N;-\rangle\approx 0$ and we see no or nearly no interference fringes in the probability density. For the isotropic 2DHO, the state at the vortex limit is perfectly Gaussian and completely non-interfering, though this coincidence of \textit{minimal} quantum interference with \textit{none at all} is specific to the isotropic case, as we shall discuss further in the next section. As we approach the standing wave limit, the eccentricity of the Lissajous ellipse increases drawing non-trivial counter-flowing currents into closer contact leading to more and more substantial cancellation. In this situation the branches of the state vector overlap to a non-negligible degree, $\langle\zeta,N;+|\zeta,N;-\rangle\neq 0$, indicating that the state of the system involves interference between these two distinct states; this shows up in the probability density function in the form of the observed interference fringes. At the standing wave limit, we see that the cancellation of the current is complete and quantum interference fringes reach their maximum visibility. The lesson here is this:\textbf{ cancellation of the probability current in configuration space is the result of quantum interference.} In other words, there is an essential trade-off between the laminar flow of probability current and the emergence of quantum interference.  

We conclude this section by commenting on one more feature of the LCS for the isotropic 2DHO. Using the same basic argument as in the previous paragraph it is clear upon inspection of Fig. (\ref{fig1:IsoVortexLim}) that it is at the vortex limit that the counter-flowing branches of the state vector are the closest to being truly orthogonal. As we move to intermediate vortex states, such as the one represented in Fig.(\ref{fig4:IsoVortexIntermediate}) we see the first emergence of non-trivial structure in form of pronounced peaks the probability density near the classical turning points along the major axis of the Lissajous ellipse. The origin of these peaks is now clear. Nontrivial contributions to the overlap, $\langle\zeta,N;+|\zeta,N;-\rangle$, between branches of the state vector will first accrue from locations in which probability current begins to \textit{double back} on itself, i.e., near precisely those turning points. The resulting peaks in the probability density are the first recognizable features of an interference pattern, and in this case happen to represent regions of constructive interference. As we can easily see, these constructive interference fringes persist all the way through the standing wave limit.  

\section{The anisotropic oscillator}

In the more general case of the anisotropic harmonic oscillator with commensurate frequencies $\omega_x=q\omega$ and $\omega_y=p\omega $   where $p$ and $q$  are coprime numbers, the Hamiltonian is given by

\begin{equation}
H_{aniso}=\hbar\omega\left[q\left(a_x^\dagger a_x+1/2\right)+p\left(a_y^\dagger a_y+1/2\right)\right]
    \label{eq:Haniso}
\end{equation}
which, as a subset of all eigenstates, has the accidentally degenerate eigenstates $|pK\rangle_x|q\left(N-K\right)\rangle_y$ $K=0,1,2,...,N$  with energy eigenvalues $\hbar\omega\left[Npq+\left(p+q\right)/2\right]$.  Note that the degenerate states in this case satisfy the eigenvalue problem
\begin{equation}
\left(qa_x^\dagger a_x+pa_y^\dagger a_y\right)|pK\rangle_x|q\left(N-K\right)\rangle_y=Npq|pK\rangle_x|q\left(N-K\right)\rangle_y
    \label{eq:anisoeigen}
\end{equation}
where there are again $N+1$  degenerate states indicating an underlying connection to the group SU(2).  The connection between these accidental degeneracies and SU(2) has been discussed by many authors  \cite{dulock_degeneracy_1965}, \cite{louck_canonical_1973}, \cite{glauber_quantum_1963}. In fact, the authors of Ref. \cite{dulock_degeneracy_1965} introduce generalized Boson operators for the anisotropic case, which were then used by Kumar and Dutta-Roy \cite{kumar_commensurate_2008} to construct a new set of operators that close on the su(2) Lie algebra. These latter authors used this construction in order to justify the SU(2)-like coherent states that appear in the paper by Cheng and Huang [1]. However, in what follows with respect to the cases of commensurate anisotropic frequencies, we ignore the above-mentioned group theoretical aspects of the situation, apart from the degeneracies themselves, which we use to construct the relevant projection operators. We are again projecting out of a product of ordinary coherent states associated with the $x$ and $y$ degrees of freedom, states that follow the classical motion, onto sets of degenerate states. In contrast to the isotropic case, the projected states are not of the usual form SU(2) coherent states.   

The projection operators on degenerate subspaces in this case are given by 

\begin{equation}
\Pi_{Npq}=\sum_{K=0}^N|pK\rangle_x|q\left(N-K\right)\rangle_{yx}\langle pK|_{y}\langle q\left(N-K\right)|
    \label{eq:anisoprojop1}
\end{equation}
which we use to perform a projection out of $|\alpha\rangle_x|\beta\rangle_y$  to obtain 

\begin{equation}
\Pi_{Npq}|\alpha\rangle_x|\beta\rangle_y\propto\sum_{K=0}^N\frac{\alpha^{pK}\beta^{q\left(N-K\right)}}{\sqrt{\left(pK\right)!\left(qN-qK\right)!}}|pK\rangle_x|q\left(N-K\right)\rangle_y.
    \label{eq:anisoprojection}
\end{equation}
The normalized form of this state is 
\begin{equation}
|\alpha,\beta,N,p,q\rangle=\mathcal{N}_{Npq}\sum_{K=0}^N\left[\frac{\left(qN\right)!}{\left(pK\right)!\left(qN-qK\right)!}\right]^{1/2}\left(\frac{\alpha^p}{\beta^q}\right)^K|pK\rangle_x|q\left(N-K\right)\rangle_y
    \label{eq:anisonormprojstate}
\end{equation}
where 

\begin{equation}
\mathcal{N}_{Npq}=\left[\sum_{L=0}^N\frac{\left(qN\right)!}{\left(pL\right)!\left(qN-qL\right)!}\left|\frac{\alpha^p}{\beta^q}\right|^{2L}\right]^{-1/2}
    \label{eq:anisonormfactor}
\end{equation}
For the case $p=q=1$  we recover the SU(2) coherent state for a two-dimensional harmonic oscillator or for a two-mode field. If we set, for a given $p$ and $q$, $\zeta_=\alpha^p/\beta^q$ then we can write our states as 
\begin{equation}
|\zeta,N,p,q\rangle=\mathcal{N}_{Npq}\sum_{K=0}^N\left[\frac{\left(qN\right)!}{\left(pK\right)!\left(qN-qK\right)!}\right]^{1/2}\zeta^K|pK\rangle_x|q\left(N-K\right)\rangle_y
    \label{eq:anisonormprojstatezetapq}
\end{equation}
with

\begin{equation}
\mathcal{N}_{Npq}=\left[\sum_{L=0}^N\frac{\left(qN\right)!}{\left(pL\right)!\left(qN-qL\right)!}\left|\zeta\right|^{2L}\right]^{-1/2}.
    \label{eq:anisonormfactorzetapq}
\end{equation}
Note that these states satisfy the relation 

\begin{equation}
\left(qa_x^\dagger a_x+pa_y^\dagger a_y\right)|\zeta,N,p,q\rangle=Npq|\zeta,N,p,q\rangle
    \label{eq:eigenaniso}
\end{equation}
In Appendix B we discuss an alternative method of arriving at these states by algebraic means.




The wave function for the general case will now be given by

\begin{equation}
\Psi_{N}^{\left(p,q\right)}\left(x,y,\zeta\right)=\mathcal{N}_{Npq}\sum_{K=0}^{N}\sqrt{\frac{\left(qN\right)!}{\left(pK\right)!\left(qN-qK\right)!}}\zeta^{K}\psi_{pK}^{\left(q\right)}\left(x\right)\psi_{q\left(N-K\right)}^{\left(p\right)}\left(y\right),
    \label{eq:anisoWF}
\end{equation}
where
\begin{equation}
\psi_{pK}^{\left(q\right)}\left(x\right)=\frac{1}{\sqrt{2^{pK}\left(pK\right)!}}\left(\frac{mq\omega}{\pi\hbar}\right)^{1/4}\exp{\left[-\frac{mq\omega}{2\hbar}x^2\right]}H_{pK}\left(x\sqrt{mq\omega/\hbar}\right)
    \label{eq:anisoxwf}
\end{equation}
and
\begin{equation}
\psi_{q\left(N-K\right)}^{\left(p\right)}\left(y\right)=\frac{1}{\sqrt{2^{q\left(N-K\right)}\left[q\left(N-K\right)\right]!}}\left(\frac{mp\omega}{\pi\hbar}\right)^{1/4}\exp{\left[-\frac{mp\omega}{2\hbar}y^2\right]}H_{pK}\left(y\sqrt{mp\omega/\hbar}\right).
    \label{eq:anisoywf}
\end{equation}
The probability distribution and current density are again given as in Eqs. (\ref{eq:isoprobdens}) and (\ref{eq:isocurrdens}). 

The polar form for the complex amplitude of the anisotropic Lissajous coherent state is $\zeta=|\zeta|e^{i\vartheta}$, where, in terms of the original ordinary coherent state amplitudes, $|\zeta|=|\alpha|^p/|\beta|^q$ and $\vartheta=p\theta_x-q\theta_y$ and where $\theta_x$ and $\theta_y$ are the phase angles for the amplitudes of the ordinary coherent states $|\alpha\rangle_x$ and $|\beta\rangle_y$, respectively. The phase condition relevant to the characterization of anisotropic standing wave and vortex states becomes

\begin{equation}
\vartheta=\begin{cases}
n\pi & \text{standing wave}\\
\left(n+1/2\right)\pi &\text{vortex limit}
\end{cases}
    \label{eq:anisophasecond}
\end{equation}
for $n=0,\pm 1,\pm 2,...$. Exactly as in the isotropic case, for complex amplitudes having phase angles other than those given by Eq. (\ref{eq:anisophasecond}), the state is generically an intermediate vortex state exhibiting neither maximal nor minimal quantum interference.

\begin{figure}
\centering
       \includegraphics[width=1.0\textwidth]{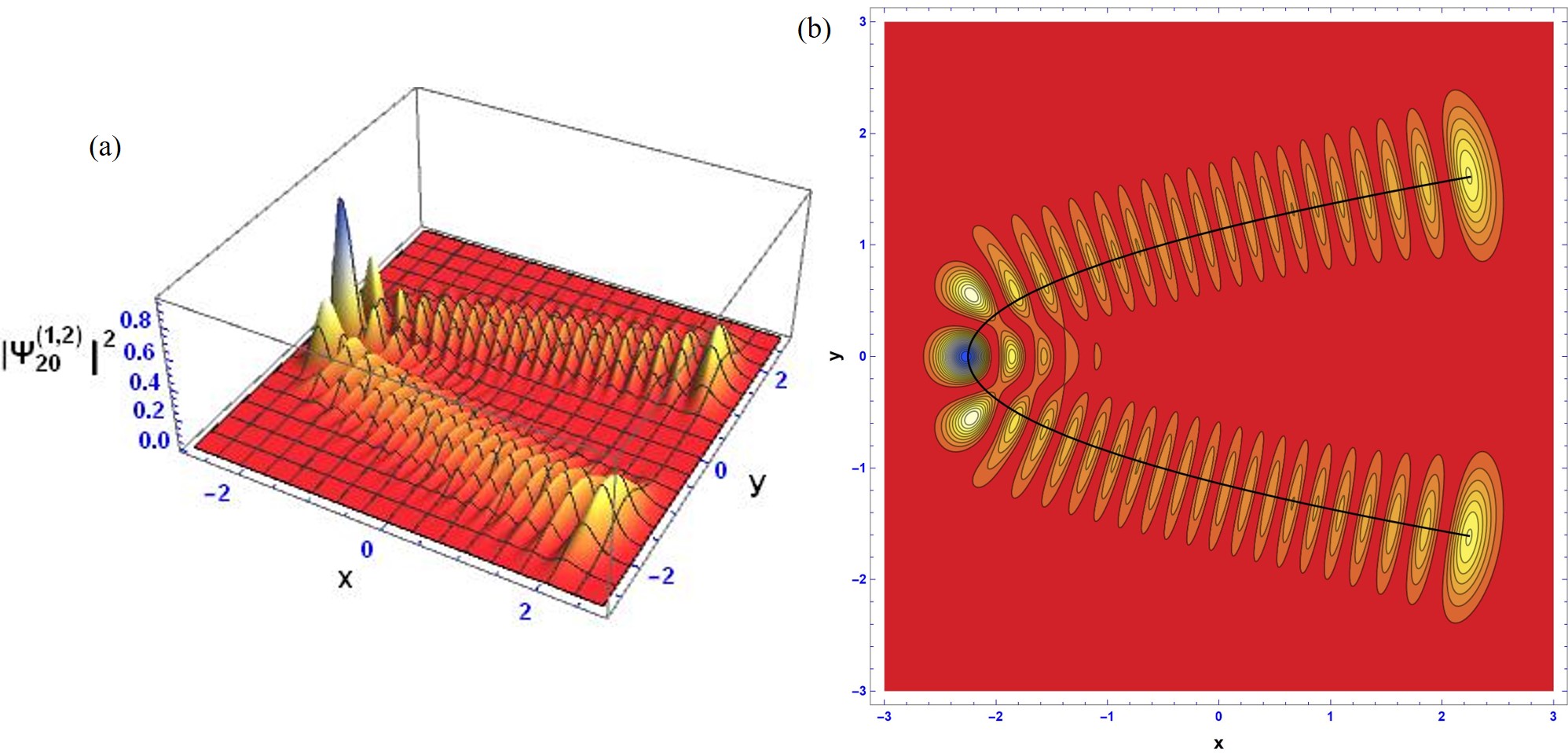}
 \caption{(a) Surface and (b) contour plot of the configuration space probability density for a LCS of the anisotropic 2DHO for which $p=1$ and $q=2$. The black line inset into (b) is the corresponding classical Lissajous trajectory.   }
\label{fig6:p=1,q=2_standing wave}
\end{figure}

To demonstrate the important features of the LCS for the anisotropic 2DHO we plot in Figs. (\ref{fig6:p=1,q=2_standing wave}) through (\ref{fig8:p=1,q=3_p=3_q=5_standing waves}) the probability densities for the state at the standing wave limit for $\left(p=1,q=2\right)$, $\left(p=2,q=3\right)$,$\left(p=1,q=3\right)$, and $\left(p=3,q=5\right)$, respectively. In each case, the states exhibit the behaviors expected at the standing wave limit; the current densities vanish identically, the quantum interference fringes attain maximum visibility, and the states are clearly localized along the classical Lissajous trajectory appropriate to each set of coprime commensurate frequencies. Further, the wavefunction phase (not shown) for each case is independent of position.

\begin{figure}
\centering
       \includegraphics[width=1.0\textwidth]{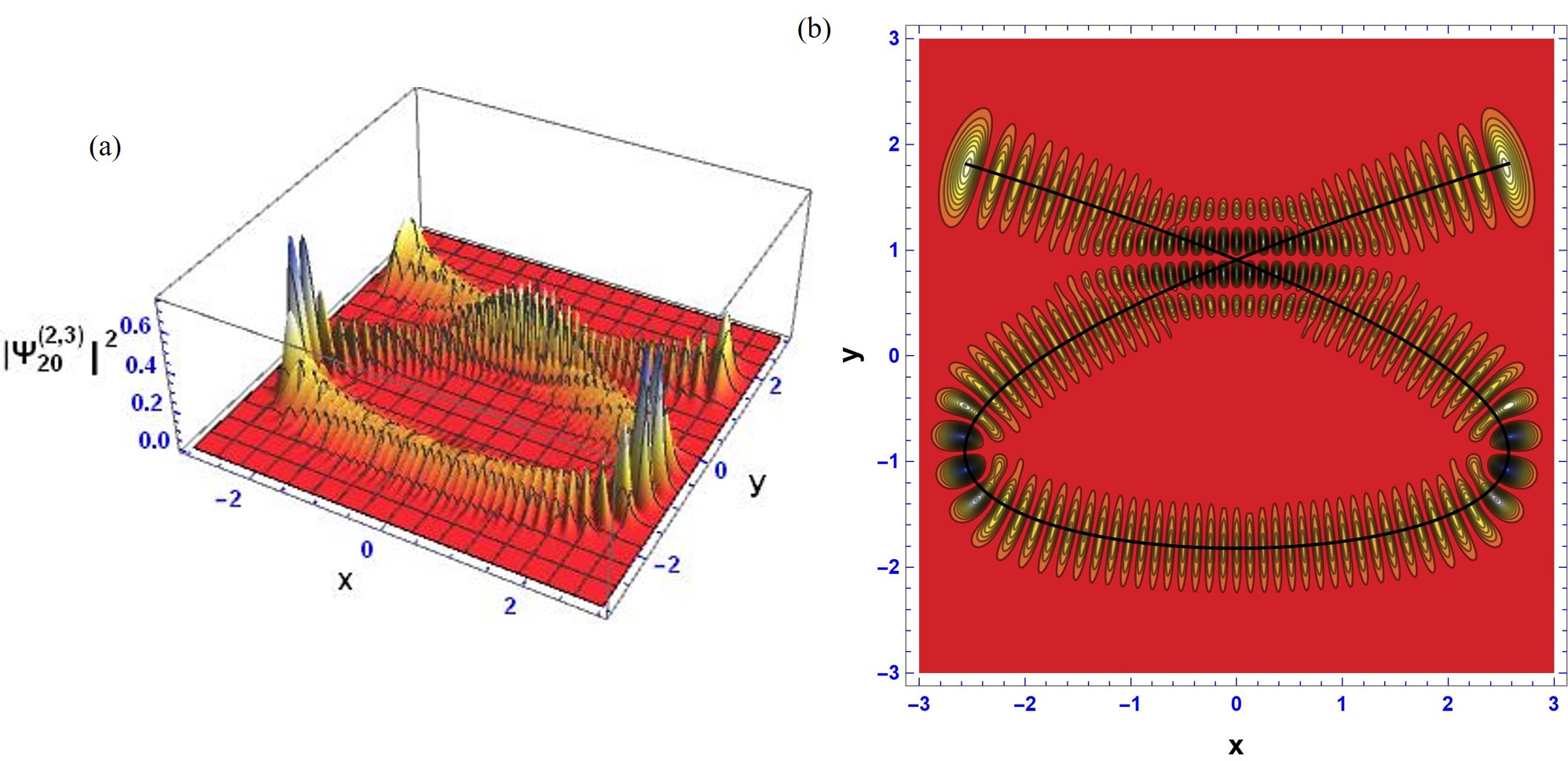}
 \caption{(a) Surface and (b) contour plot of the configuration space probability density for a LCS of the anisotropic 2DHO for which $p=2$ and $q=3$. The black line inset into (b) is the corresponding classical Lissajous trajectory. Here, as in Fig. (\ref{fig6:p=1,q=2_standing wave}), the probability current density vanishes identically yielding quantum interference fringes of maximum visibility in the probability density.   }
\label{fig7:p=2,q=3_standing wave}
\end{figure}
\begin{figure}
\centering
       \includegraphics[width=1.0\textwidth]{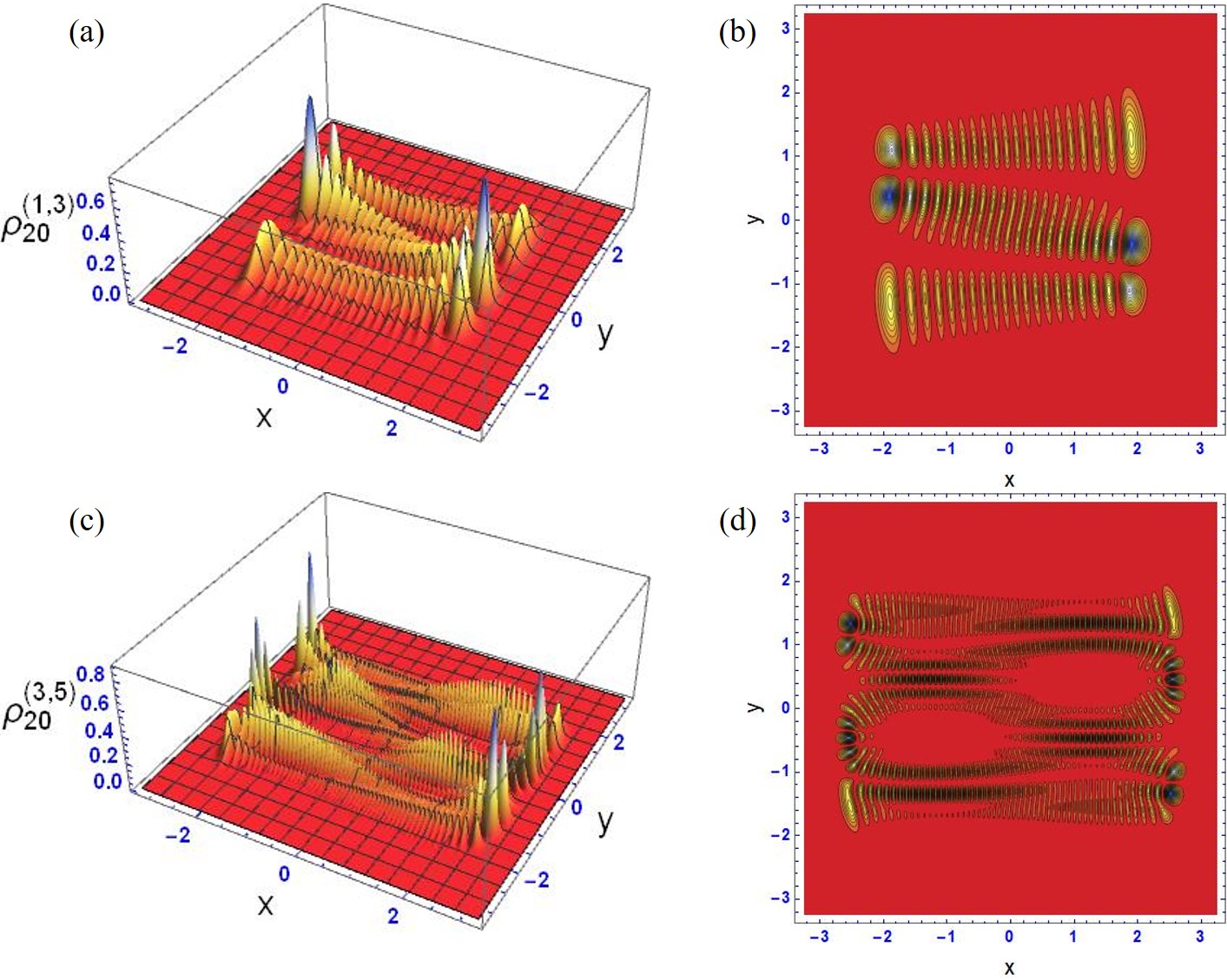}
 \caption{Surface and contour plots of the LCS at the standing wave limit for $p=1, q=3$ (a) and (b) and $p=3,q=5$ (c) and (d). The classical Lissajous trajectories are omitted here for visual clarity of the figures.  }
\label{fig8:p=1,q=3_p=3_q=5_standing waves}
\end{figure}

To illustrate the nature of vortex states of the anisotropic 2DHO, we show in Fig. (\ref{fig9:p=1,q=2_VortexLim_prob_current_classical}) the probability and probability current densities overlain with the classical Lissajous trajectory for the state having $p=1$ and $q=2$ along with the phase of the configuration space wavefunction for the LCS at the vortex limit. In addition to localization around the corresponding classical Lissajous trajectory this figure exposes several important features of the vortex states of the anisotropic 2DHO. First, it is clear that for the anisotropic 2DHO just as for the isotropic one the LCS at the vortex limit of the quantum system corresponds to the classical Lissajous trajectory that is symmetric in two orthogonal directions; for the parameters we have chosen here those are the $x$ and $y$ directions, but that would not have to be the case. In any case this symmetry property is expected in light of the Ehrenfest Theorem, according to which the phase difference between the complex amplitude of the LCS is equal to the phase difference between the corresponding classical oscillators. Second, unlike the case for the vortex limit in the isotropic case, quantum interference fringes never fully vanish for the LCS at the vortex limit of the 2DHO. The reason for this is clear upon careful inspection of Fig. (\ref{fig9:p=1,q=2_VortexLim_prob_current_classical}). Even at the vortex limit, there are regions where otherwise non-negligible counter-flowing probability currents overlap and cancel to some degree. In these regions we observe non-trivial interference fringes, see for example the regions of the plot around $\left(x\approx\pm1.8,y\approx\pm1\right)$, especially in contrast with Fig. (\ref{fig1:IsoVortexLim}b), and for that matter, in comparison with the basic situation in Fig. (\ref{fig5:IsoVortex_Pi_over_8}b) for the isotropic 2DHO.   

We can now appreciate the full Lissajous nature of the states we have derived via projection out of a product of coherent states onto a degenerate subspace of the 2DHO. These states are clearly localized along the corresponding classical Lissajous trajectory in the sense of support of the probability and the probability current densities; this is in fact the property that inspired the name for this class of quantum states of the 2DHO. We have discussed and demonstrated this correspondence in connection with the orientations and shapes of several examples of 2DHOs having commensurate frequencies. Further, in the quantum mechanical case, we have established the close connection between the laminar flow of probability current and the emergence (or disappearance) of a quantum interference pattern in the probability density for the state. The underlying structure connecting these things is the phase of the configuration space wavefunction, which has trivial geometric singularities that serve to delineate regions of counter-flowing current. Wherever the current is otherwise non-trivial at the location of such a singularity, there is a complete cancellation of it, indicating the presence of interference fringes. For the state represented in Fig. (\ref{fig9:p=1,q=2_VortexLim_prob_current_classical}), these occur along arcs on which the current density completely flows back upon itself. Near the classical turning points along the $x$ direction, these cancellations involve significant levels of current, resulting in the presence of visible interference fringes in those regions. In principle, knowing the distribution of these geometric singularities would allow one to construct a qualitative picture of the quantum interference properties of a LCS. 
\begin{figure}
\centering
       \includegraphics[width=1.0\textwidth]{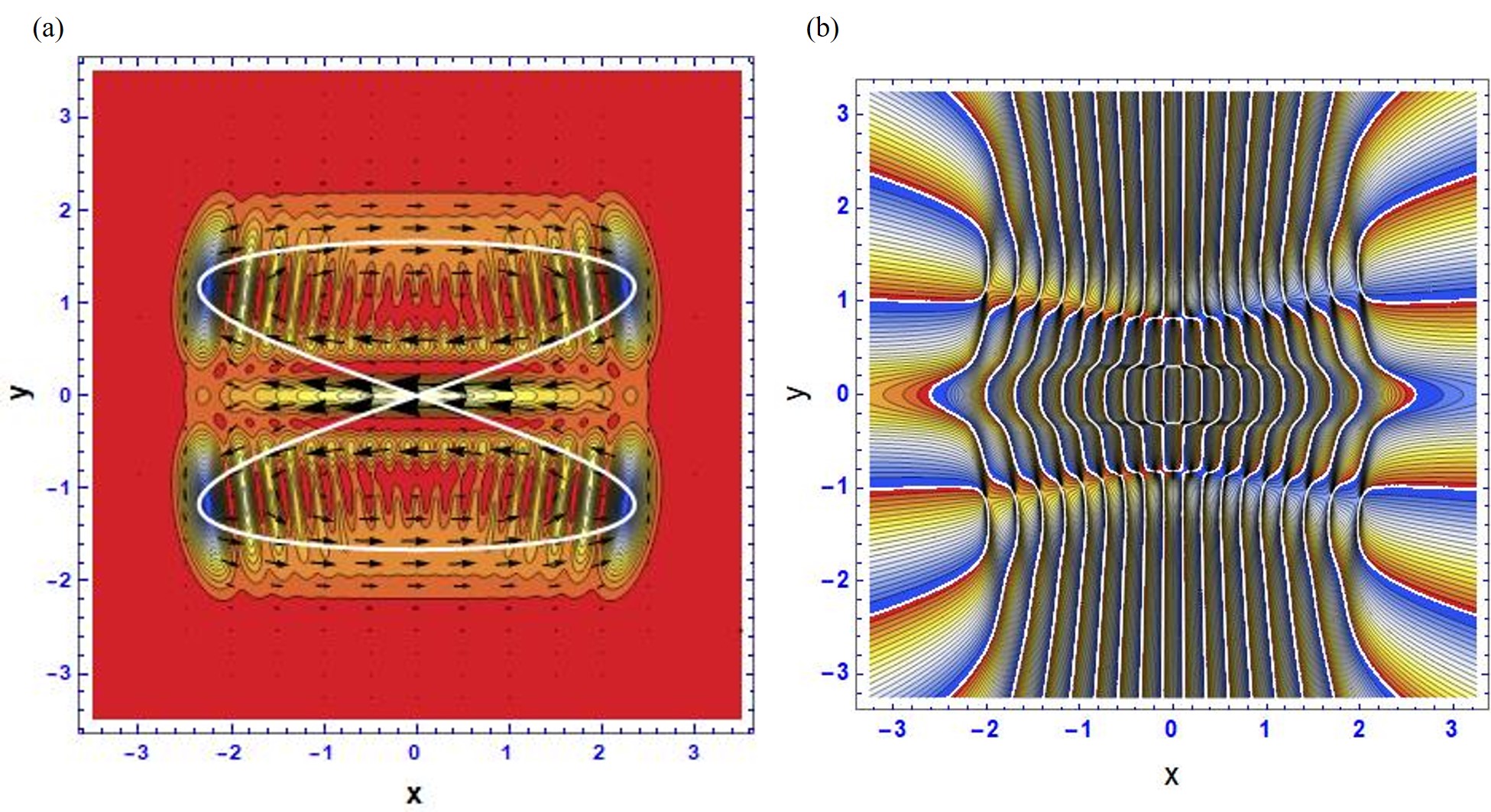}
 \caption{(a) Contour plots of the probability density overlain with the vector field plot for the probability current density (black arrows) and the classical Lissajous trajectory (white) for a state at the vortex limit of an anisotropic 2DHO having $p=1$ and $q=2$. (b) Contour plot of the phase of the wavefunction for this state. Trivial arithmetic jumps in the phase appear in white, and trivial geometric singularities associated with the ambiguous sense of circulation for counter-flowing currents show up at inflection points along the rays characterizing the jump discontinuities. Here again, these are purely mathematical artifacts. }
\label{fig9:p=1,q=2_VortexLim_prob_current_classical}
\end{figure}

\begin{figure}
\centering
       \includegraphics[width=1.0\textwidth]{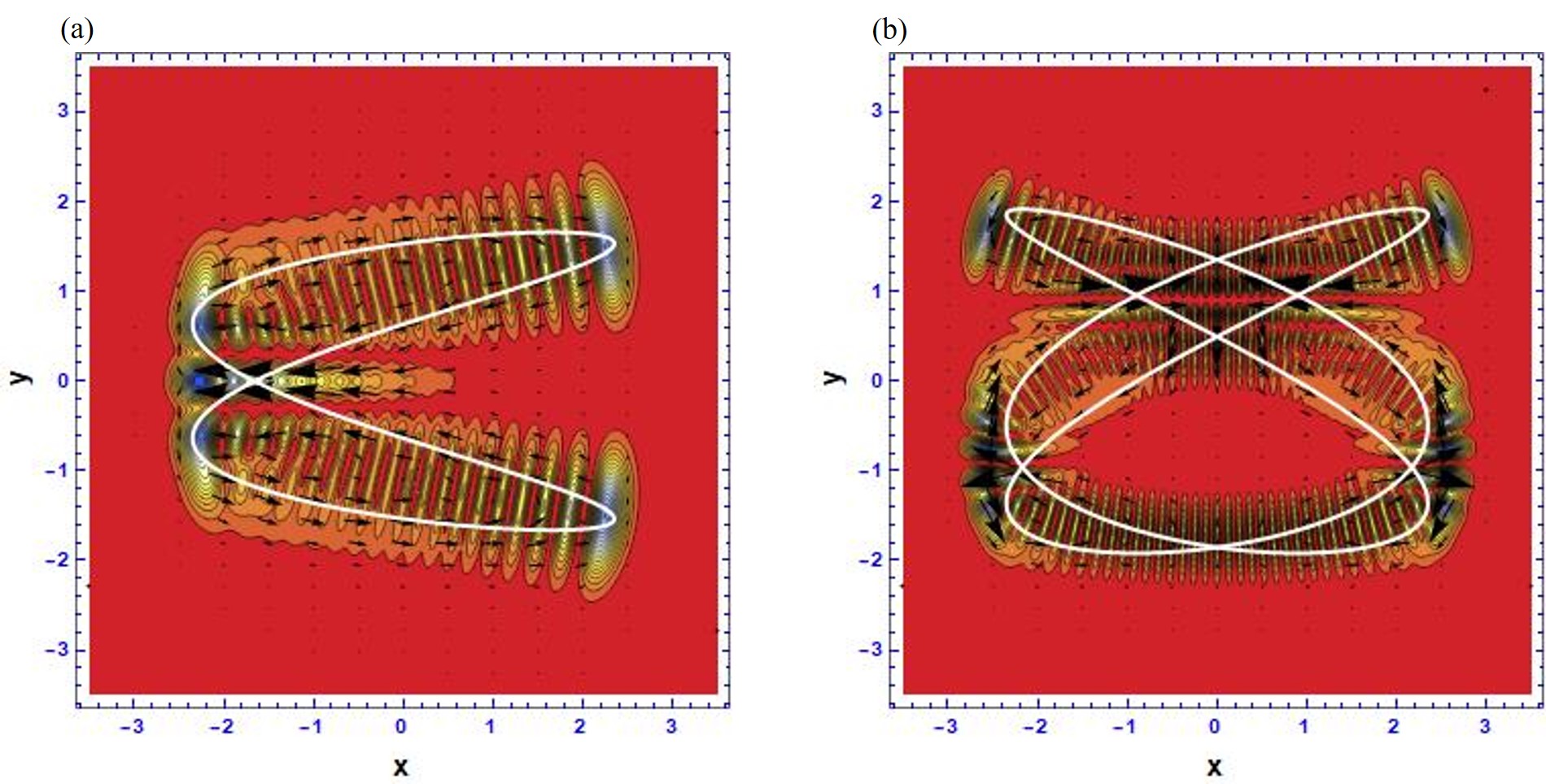}
 \caption{Contour plots of the probability density overlain with the vector field plot for the probability current density (black arrows) and the classical Lissajous trajectory (white) for intermediate vortex states of an anisotropic 2DHO having (a) $p=1$ and $q=2$ and (b) $p=2$ and $q=3$. The phase of the complex amplitude of each LCS state vector is set at $\pi/4$, making  these examples the intermediate vortex states halfway between the standing wave limit and the vortex limit for each case. }
\label{fig10:AnisoIntVortex}
\end{figure}

To complete our analysis of the LCS, we plot in Fig. (\ref{fig10:AnisoIntVortex}) the same quantities as in Fig. (\ref{fig9:p=1,q=2_VortexLim_prob_current_classical}a), but for intermediate vortex states of the anisotropic 2DHO having $p=1, q=2$ and $p=2, q=3$. All of the features discussed in detail so far are apparent in these plots. Now, we see also the expected asymmetry in the shape of the Lissajous figure, which is, again, expected as a result of the Ehrenfest Theorem. Although we do not display it here, a corresponding asymmetry is exhibited in the phase of the wavefunction for each of these states and, in particular, in the distribution of its geometric singularities and jump discontinuities.   

\section{Summary and outlook}
In this paper we have addressed the issue of obtaining stationary states of an anisotropic two-dimensional harmonic oscillator of coprime frequencies, that are concentrated on the corresponding classical orbits, namely, the Lissajous figures, by mathematically performing projections onto sets of degenerate states out of products of ordinary coherent states,   for the independent motions along the $x$ and $y$ directions as indicated. The basic idea is that we start from quantum states known to follow the classical motion of any two-dimensional harmonic oscillator (whether the frequencies are coprime or not). Furthermore, the idea of constructing our states for the anisotropic cases follows as an extension of the fact that, for the isotropic case, the resulting projections yield the SU(2) coherent states, results obtained by De Bièvre \cite{bievre_oscillator_1992} and Pollet et al. \cite{pollett_elliptic_1995} by other means. The states we obtain for the coprime anisotropic cases are not SU(2) coherent states. In fact, they are very different than the corresponding states constructed by Chen and Huang \cite{chen_vortex_2003}. 

The states we discuss are new coherent states in that they satisfy certain algebraic relations, and can be derived iteratively from those relations, as described in Appendix B. Furthermore, these states have been shown to resolve unity on the finite dimensional Hilbert space over the relevant degenerate states, as shown in Appendix A. In fact, these states are overcomplete in these spaces, just as should be the case for states that can legitimately be called “coherent states.” This is so because the completeness relations obtained are the result of applying the relevant projection operators to the product of the integrals resolving unity for the coherent states associated with the $x$ and $y$ directions. For the isotropic case we have shown that the resulting completeness relation can be rendered into the usual form of the resolution of unity expected for the SU(2) coherent states by a transformation of variables. 

In addition, we have clarified the nature of singularities present in the phase of the configuration space wavefunction for the LCS. In particular, we has shown that the only singularities that occur in the phase are purely geometric in nature having no physical significance. Other apparent jump discontinuities are not really present at all. Rather, they turn out to be arithmetic artifacts arising from our quantification of the phase$\mod 2\pi.$

Finally, we have formulated on robust theoretical ground the important trade-off between the laminar flow of steady probability current density and the emergence of a sharp quantum interference pattern in the probability density. In so doing, we have derived constraints on the phases of the complex amplitudes for the state vectors that unambiguously define the vortex and standing wave limits, thus providing a clear and quantifiable meaning for the designation of a LCS as a vortex state.

An interesting class of LCS having no classical analogs are those that occur when we take $p=rp_0$ and $q=rq_0$ for some integer $r>1$ and where $p_0$ and $q_0$ are coprime. We refer to these types of states as \textit{Higher Harmonic} LCS (HHLCS), and they occur in two sub-classes. Those for which $p_0=q_0=1$, we call isotropic HHLCS and another for which $p_0\neq q_0$ anisotropic HHLCS. While these types of states have been presented in the literature \cite{chen_vortex_2003}, we will examine in an upcoming paper the specific forms of these resulting from projection out of ordinary coherent states. An advantage of approaching the problem using this organic approach has been, among other things, the clarity with which we can expose and understand the important, and often very deep, quantum mechanical properties of the resulting states. This carries over into the higher harmonic versions of the LCS, notably resulting in an important connection between the HHLCS and coherent superpositions of SU(2) coherent states and their generalizations as we have presented them here. Further, developing a theory of LCS via projection could have important utility in the search for systems in which they arise experimentally, which is another avenue of our ongoing work that we will present elsewhere. 

\section*{Acknowledgments}
The views expressed are those of the authors and do not reflect the official guidance or position of the United States Government, the Department of Defense, or of the United States Air Force. The appearance of external hyperlinks does not constitute endorsement by the United States Department of Defense (DoD) of the linked websites, or of the information, products, or services contained therein. The DoD does not exercise any editorial, security, or other control over the information you may find at these locations.

\section*{Appendix A: Completeness relations}

Completeness, in fact, over-completeness, is an important characteristic of all coherent states. Coherent states are expected to represent the identity operator in a continuous basis, involving integrations over the complex parameters that define the states. Here we show that the states discussed in this paper are legitimate coherent states in the sense that they can resolve the relevant identity operators in the spaces over which the states are defined.

We start by writing down the product of completeness relations for the two sets of ordinary coherent states: 

\begin{equation}
\int\int\frac{d^2\alpha d^2\beta}{\pi^2}|\alpha,\beta\rangle\langle\alpha,\beta|=I_xI_y
\label{eq:GlauberCompleteness}
\end{equation}
where $|\alpha,\beta\rangle\equiv|\alpha\rangle_x|\beta\rangle_y$  and where $I_i=\sum_{n=0}^{\infty}|n\rangle_{ii}\langle n|$ for $i=x,y$.  The projection operator for the general case of the states discussed above is 

\begin{equation}
\Pi_{Npq}=\sum_{K=0}^N|pK\rangle_x|q\left(N-K\right)\rangle_{yx}\langle pK|_y\langle q\left(N-K\right)|.
\label{eq:anisoprojop2}
\end{equation}
We apply this operator to Eq. (\ref{eq:GlauberCompleteness}) from both sides to obtain
\begin{equation}
\int\int\frac{d^2\alpha d^2\beta}{\pi^2}\Pi_{Npq}|\alpha,\beta\rangle\langle\alpha,\beta|\Pi_{Npq}=\Pi_{Npq}I_xI_y\Pi_{Npq}=\Pi_{Npq}.
\label{eq:anisoCompleteness1}
\end{equation}
It is a straightforward, though tedious, exercise to show, upon doing the integrals, that the left-hand side of Eq. (\ref{eq:anisoCompleteness1}) yields the right-hand side of that equation, as it must. The left-hand side of this equation can be rewritten such that we obtain 

\begin{equation}
\int\int\frac{d^2\alpha d^2\beta}{\pi^2}\frac{e^{-\left(|\alpha|^2+|\beta|^2\right)}}{\left[\mathcal{N}_{Npq}\left(\alpha,\beta\right)\right]^2}|\alpha,\beta,N,p,q\rangle\langle\alpha,\beta,N,p,q|=\Pi_{Npq}
\label{eq:anisoCompleteness2}
\end{equation}
where

\begin{equation}
|\alpha,\beta,N,p,q\rangle=\mathcal{N}_{Npq}\left(\alpha,\beta\right)\sum_{K=0}^N\frac{\alpha^{pK}\beta^{q\left(N-K\right)}}{\sqrt{\left(pK\right)!\left(qN-qK\right)!}}|pK\rangle_x|q\left(N-K\right)\rangle_y
\label{eq:anisostatevec}
\end{equation}
and where

\begin{equation}
\mathcal{N}_{Npq}\left(\alpha,\beta\right)=\left[\sum_{K=0}^N\frac{|\alpha|^{2pK}|\beta|^{2q\left(N-K\right)}}{\left(pK\right)!\left(qN-qK\right)!}\right]^{-1/2}
\label{eq:anisonorm}
\end{equation}
Equation (\ref{eq:anisoCompleteness2}) constitutes the formal completeness relation for our states with arbitrary $N$, $p$, and $q$.

For the special case where $p=q=1$   we have that 

\begin{equation}
\int\int\frac{d^2\alpha d^2\beta}{\pi^2}\Pi_N|\alpha,\beta\rangle\langle\alpha,\beta|\Pi_N=\Pi_N
\label{eq:isoCompleteness0}
\end{equation}
which can be used directly to arrive at

\begin{equation}
\int\int\frac{d^2\alpha d^2\beta}{\pi^2}e^{-\left(|\alpha|^2+|\beta|^2\right)}\frac{|\beta|^{2N}}{N!}\left(1+\left|\frac{\alpha}{\beta}\right|^2\right)^N|\frac{\alpha}{\beta},N\rangle\langle\frac{\alpha}{\beta},N|=\Pi_N
\label{eq:isoCompleteness}
\end{equation}
where

\begin{equation}
|\frac{\alpha}{\beta},N\rangle\equiv\left(1+|\frac{\alpha}{\beta}|^2\right)^{-N/2}\sum_{K=0}^N\binom{N}{K}^{1/2}\left(\frac{\alpha}{\beta}\right)^K|K\rangle_x|N-K\rangle_y
\label{eq:SU2CSalphabeta}
\end{equation}
is the SU(2) coherent state that we previously given in Eq.(\ref{eq:NormSU2CS}) if we set $\alpha/\beta=\zeta$.  

We next perform a transformation to a new set of variables. To set the stage for this we first note that

\begin{equation}
d^2\alpha=d\alpha_x d\alpha_y, d^2\beta=d\beta_x d\beta_y
\label{eq:dalphadbeta}
\end{equation}
where we have set $\alpha=\alpha_x+i\alpha_y$ and $\beta=\beta_x+i\beta_y $.  We now write the new variable $\zeta$  as 

\begin{equation}
\zeta=\frac{\alpha}{\beta}=\frac{\alpha\beta^*}{|\beta|^2}=\frac{\alpha_x\beta_x+\alpha_y\beta_y}{|\beta|^2}+i\frac{\alpha_y\beta_x-\alpha_x\beta_y}{|\beta|^2}=\zeta_x+i\zeta_y,
\label{eq:dzeta}
\end{equation}
where

\begin{equation}
\zeta_x=\frac{\alpha_x\beta_x+\alpha_y\beta_y}{|\beta|^2}, \zeta_y=\frac{\alpha_y\beta_x-\alpha_x\beta_y}{|\beta|^2}.
\label{eq:zetaxy}
\end{equation}
We now let $w=\beta$  such that $w_x=\beta_x$ and $ w_y=\beta_y $  

This allows us to write 

\begin{equation}
\int\int d^2\alpha d^2\beta=\int\int d^2\zeta d^2w \left|\begin{array}{cccc}\cfrac{\partial \zeta_x}{\partial \alpha_x}&\cfrac{\partial \zeta_x}{\partial \alpha_y}&\cfrac{\partial \zeta_x}{\partial \beta_x}&\cfrac{\partial \zeta_x}{\partial \beta_y} \\ \cfrac{\partial \zeta_y}{\partial \alpha_x}&\cfrac{\partial \zeta_y}{\partial \alpha_y}&\cfrac{\partial \zeta_y}{\partial \beta_x}&\cfrac{\partial \zeta_y}{\partial \beta_y}\\ \cfrac{\partial w_x}{\partial \alpha_x}&\cfrac{\partial w_x}{\partial \alpha_y}&\cfrac{\partial w_x}{\partial \beta_x}&\cfrac{\partial w_x}{\partial \beta_y}\\\cfrac{\partial w_y}{\partial \alpha_x}&\cfrac{\partial w_y}{\partial \alpha_y}&\cfrac{\partial w_y}{\partial \beta_x}&\cfrac{\partial w_y}{\partial \beta_y}\end{array}\right|^{-1}.
\label{eq:Jacobian}
\end{equation}
The Jacobian has the value $1/|w|^2=1/|\beta|^2$  such that 

\begin{equation}
\int\int d^2\alpha d^2\beta=\int\int d^2\zeta d^2w|w|^2=\int\int d^2\zeta d^2\beta|\beta|^2.
\label{eq:transform}
\end{equation}
Equation (\ref{eq:isoCompleteness}) now becomes 

\begin{equation}
\int\int\frac{d^2\alpha d^2\beta}{\pi^2}e^{-\beta|^2\left(1+|\zeta|^2\right)}\frac{|\beta|^{2N+2}}{N!}\left(1+|\zeta|^2\right)^N|\zeta,N\rangle\langle\zeta,N|=\Pi_N,
\label{eq:isobetazeta}
\end{equation}
It remains for us to perform the integral over $\beta$   which we do by setting $\beta=r e^{i\phi}$  such that 

\begin{equation}
\int d^2\beta e^{-\beta|^2\left(1+|\zeta|^2\right)}|\beta|^{2N+2}=\int_{0}^{\infty}\int_{0}^{2\pi}rdrd\phi e^{-r^2\left(1+|\zeta|^2\right)}r^{2N+2}=2\pi\int_{0}^{\infty}dr e^{-r^2\left(1+|\zeta|^2\right)}r^{2N+3},
\label{eq:betapolar}
\end{equation}
and, making efficient us of the Gamma function,

\begin{equation}
\int d^2\beta e^{-\beta|^2\left(1+|\zeta|^2\right)}|\beta|^{2N+2}=2\pi\int_{0}^{\infty}dr e^{-r^2\left(1+|\zeta|^2\right)}r^{2N+3}=\pi\frac{\left(N+1\right)!}{\left(1+|\zeta|^2\right)^{N+2}}.
\label{eq:betaresult}
\end{equation}
Putting this all together we find that Eq.(\ref{eq:isoCompleteness}) can now be written as 

\begin{equation}
\frac{N+1}{\pi}\int  \frac{d^2\zeta}{\left(1+|\zeta|^2\right)^2}|\zeta,N\rangle\langle\zeta,N|=\Pi_N,
\label{eq:SU2resunity}
\end{equation}
which is the correct form for the resolution of unity in terms of the SU(2) coherent states $|\zeta,N\rangle$  in the context of the isotropic two-dimensional harmonic oscillator for a space of $N+1$   degenerate states. 

We point out here that, for the case where $p=q=1$   in contrast to the paper of Moran and Hussin \cite{moran_coherent_2019}, we have been able to derive the proper expression for the SU(2) completeness relation organically from the action of the projection operators on the product of the two ordinary coherent state completeness relations. No equation of the form of Eq. (\ref{eq:SU2resunity}) is presented in Ref. \cite{moran_coherent_2019}.

\section*{Appendix B: Algebraic derivation of our states}
In this appendix, we describe an alternative derivation of the states given by Eq.(\ref{eq:NormSU2CS}) . This is an algebraic method that has been used in a previous work to find “dark states” in two channel models of atomic excitation where there are competing $p$- and $q$-photon process \cite{gerry_dark_2005}.  

To get started, we first consider the problem of finding solutions to the algebraic problem appropriate to the 2D isotropic oscillator problem, which is

\begin{equation}
\left(a_x-\zeta a_y\right)|\zeta\rangle=0
\label{eq:isoalgebraic}
\end{equation}
subject to the condition that

\begin{equation}
\left(a_x^\dagger a_x+ a_y^\dagger a_y\right)|\zeta\rangle=N|\zeta\rangle, N=1,2,...,
\label{eq:isoalgebraiccond}
\end{equation}
which means that the solutions must have the form 

\begin{equation}
|\zeta\rangle=\sum_{n=0}^NC_n|n\rangle_x|N-n\rangle_y.
\label{eq:isosolnform}
\end{equation}
Substitution into Eq. (\ref{eq:isoalgebraic}) leads to the recurrence relation

\begin{equation}
C_{n+1}=\zeta \sqrt{\frac{N-n}{n+1}}C_n
\label{eq:isorecursion}
\end{equation}
which can be solved to yield 

\begin{equation}
C_{n}=\zeta^n \sqrt{\frac{N!}{n!\left(N-n\right)!}}C_0
\label{eq:isorecursionsoln}
\end{equation}
which apart from normalization are the coefficients for the SU(2) coherent states. Normalization leads to $C_0=\left(1+|\zeta|^2\right)^{-N/2}$  so that 

\begin{equation}
|\zeta\rangle= \left(1+|\zeta|^2\right)^{-N/2}\sum_{n=0}^N\zeta^n\binom{N}{n}^{1/2}|n\rangle_x|N-n\rangle_y
\label{eq:isoalgebraicresult}
\end{equation}
which agrees with the result of Eq.(\ref{eq:SU2CohState}). The derivation of the SU(2) coherent state in the context of two-mode fields has previously been discussed in refs. \cite{deb_population_1993,li_influences_1995,gerry_population_1999}. 

The general case for the anisotropic oscillator with commensurate frequencies is to search for states that satisfy the equation

\begin{equation}
\left(a_x^p-\zeta a_y^q\right)|\zeta\rangle=0
\label{eq:anisoalg}
\end{equation}
subject to the condition that 

\begin{equation}
\left(qa_x^\dagger a_x+ pa_y^\dagger a_y\right)|\zeta\rangle=Npq|\zeta\rangle, N=1,2,...,
\label{eq:ANisoalgebraiccond}
\end{equation}
The solutions then must be of the form 

\begin{equation}
|\zeta\rangle=\sum_{K=0}^NC_K|pK\rangle_x|q\left(N-K\right)\rangle_y.
\label{eq:ANisosolnform}
\end{equation}
Substituting in Eq. (\ref{eq:anisoalg}) we obtain

\begin{equation}
\sum_{K=1}^NC_{K}\sqrt{\frac{\left(pK\right)!}{\left(pK-p\right)!}}|p\left(K-1\right)\rangle_x|q\left(N-K\right)\rangle_y=\zeta\sum_{K=0}^NC_{K}\sqrt{\frac{\left(qN-qK\right)!}{\left(qN-qK-q\right)!}}|pK\rangle_x|q\left(N-K-1\right)\rangle_y.
\label{eq:ANisoSUB}
\end{equation}
On the left side of Eq. (\ref{eq:ANisoSUB}) we make the replacement $K\rightarrow  K+1$ to obtain

\begin{equation}
\sum_{K=0}^{N-1}C_{K+1}\sqrt{\frac{\left(pK+p\right)!}{\left(pK\right)!}}|pK\rangle_x|q\left(N-K-1\right)\rangle_y=\zeta\sum_{K=0}^NC_{K}\sqrt{\frac{\left(qN-qK\right)!}{\left(qN-qK-q\right)!}}|pK\rangle_x|q\left(N-K-1\right)\rangle_y.
\label{eq:ANisoSUBshift}
\end{equation}
from which follows the recurrence relation 

\begin{equation}
C_{K+1}=\zeta C_K\sqrt{\frac{\left(qN-qK\right)!\left(pK\right)!}{\left(qN-qK-q\right)!\left(pK+p\right)!}}
 \label{eq:ANisorecur}   
\end{equation}
This can be iterated to yield 

\begin{equation}
C_{K}=\zeta^K \sqrt{\frac{\left(qN\right)!}{\left(pK\right)!\left(qN-qK\right)!}}C_0,
 \label{eq:ANisoiterate}   
\end{equation}
and from normalization we determine that

\begin{equation}
C_{0}=\left[\sum_{L=0}^N|\zeta|^{2L}\frac{\left(qN\right)!}{\left(pL\right)!\left(qN-qK\right)!}\right]^{-1/2}\equiv \mathcal{N}_{Npq}.
 \label{eq:ANisonorm}   
\end{equation}
Thus our normalized state is given by

\begin{equation}
|\zeta,N,p,q\rangle=\mathcal{N}_{Npq}\sum_{K=0}^N\zeta^{K}\sqrt{\frac{\left(qN\right)!}{\left(pK\right)!\left(qN-qK\right)!}}|pK\rangle_x|q\left(N-K\right)\rangle_y,
 \label{eq:ANisotate}   
\end{equation}
in agreement with Eq. (\ref{eq:anisonormprojstate}). These states were first discussed by Gerry et al \cite{gerry_population_1999} in the context of dark states in two-channel models of excitation involving competing different numbers of photons being absorbed or emitted in the two channels. Obviously, in the case where $p=q=1$ we recover the SU(2) coherent states of Eq. (\ref{eq:isoalgebraicresult}).


\end{document}